\documentclass{appolb}
\usepackage{graphicx}

\begin{document}
\title{Observational constraints on tensor perturbations in cosmological models with dynamical dark energy
\thanks{Presented at The 3rd Conference of the Polish Society on Relativity, 25-29 September 2016
Krakow, Poland at Faculty of Physics, Astronomy and Applied Computer Science, Jagiellonian University}
}
\author{O. Sergijenko
\address{Astronomical Observatory of 
Ivan Franko National University of Lviv, \\Kyryla i Methodia str., 8, Lviv, 79005, Ukraine}
}
\maketitle
\begin{abstract}
We constrain the contribution of tensor-mode perturbations with free $n_t$ in the models with dynamical dark energy with the barotropic equation of state using Planck-2015 data on CMB anisotropy, polarization and lensing, BICEP2/Keck Array data on B-mode polarization, power spectrum of galaxies from WiggleZ and SN Ia data from the JLA compilation. We also investigate the uncertainties of reconstructed potential of the scalar field dark energy.
\end{abstract}
\PACS{95.36.+x, 98.80.-k}
  
\section{Introduction}

In the last years publication of the data on B-mode polarization of CMB (e.g. \cite{bkp}) has given new opportunities to determine the contribution of tensor mode of cosmological perturbations (primordial gravitational waves).

We constrain 2 free parameters -- the tensor-to-scalar ratio $r$ and tensor spectral index $n_t$ -- jointly with the parameters of dynamical dark energy model and main cosmological ones.  We restrict consideration to the models with $-1\leq n_t\leq0$. For dark energy we adopt the model of minimally coupled classical scalar field with the barotropic equation of state described in \cite{novosyadlyj2014} (involving both quintessential and phantom subclasses).

\section{Method and data}
We use the Monte Carlo Markov chain (MCMC) method implemented in the CosmoMC code \cite{cosmomc}, assume the Universe to be spatially flat and apply for neutrinos the minimal-mass normal hierarchy of masses: a single massive eigenstate with $m_{\nu}=0.06$ eV. We apply flat priors with ranges of values [-2,-0.33] for $w_0$ and [-2,0] for $c_a^2$.

We use the following observational data: CMB TT, TE, EE angular power spectra and lensing from the Planck-2015 results \cite{planck2015}; B-mode polarization from the joint analysis of BICEP2/Keck Array and Planck (BKP) \cite{bkp}; B-mode polarization for 2 frequency channels from BICEP2/Keck Array (BK) \cite{bk}; power spectrum of galaxies from WiggleZ Dark Energy Survey \cite{wigglez}; Supernovae Ia luminosity distances from JLA compilation \cite{jla}; Hubble constant determination \cite{hst}.

\begin{figure}[htb]
\centerline{
\includegraphics[width=0.73\textwidth]{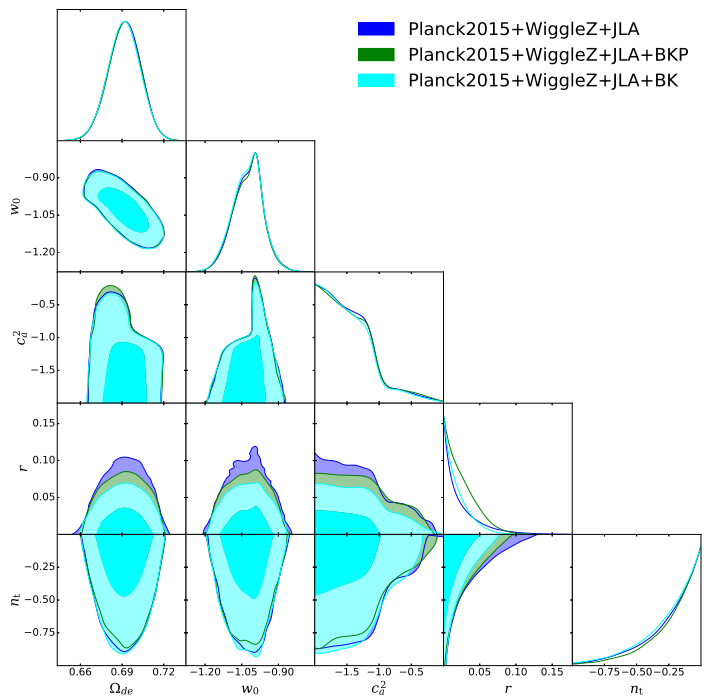}}
\caption{One-dimensional marginalized posteriors for $\Omega_{de}$, $w_0$, $c_a^2$, $r$ and $n_t$; $1\sigma$ and $2\sigma$ confidence contours from the two-dimensional marginalized posterior distributions.}
\label{combi}
\end{figure}

\section{Results and conclusions}
The results are presented in Fig. \ref{combi} and Table \ref{tab}. 

We see that the maximum of posterior for $n_t$ is at 0, where the distribution is cut by prior. Therefore, for the models with free tensor spectral index the positive values of $n_t$ should be included into analysis. The obtained 2$\sigma$ upper limit on the tensor-to-scalar ratio is significantly lowered by inclusion of BKP and especially BK data on B-mode polarization. These limits are lower than the corresponding ones obtained for the slow-roll approximation and the same cosmological model (0.118 for Planck2015+WiggleZ+JLA, 0.09 for Planck2015+WiggleZ+JLA+BKP, 0.072 for Planck2015+WiggleZ +JLA+BK).

Inclusion of the B-mode polarization data as well as using the slow-roll approximation for inflation has almost no effect on precision of determination of the other cosmological parameters.

The dark energy models corresponding to the mean values of parameters and their 1$\sigma$ and 2$\sigma$ lower confidence limits are phantom, while those with parameters at the 1$\sigma$ and 2$\sigma$ upper confidence limits are quintessence. The phantom dark energy models with mean parameters are much closer to the cosmological constant than the obtained in \cite{novosyadlyj2014} ones for all 3 datasets: their equation of state parameter evolved only from -1 to -1.023 -- -1.025 from the Big Bang up to now. This results in significantly weaker constraints on $c_a^2$ than in \cite{novosyadlyj2014}. 

\begin{table}[tbp]
\centering
 \caption{The mean values, 1$\sigma$ and 2$\sigma$ confidence limits for parameters of
cosmological models obtained from 3 observational datasets: Planck2015+WiggleZ+JLA (I), Planck2015+WiggleZ+JLA+BKP (II), Planck2015+WiggleZ+JLA+BK (III).}
\begin{tabular}{cccc}
  \hline\hline
  Parameters&(I)&(II)&(III)\\
  &mean$\pm1\sigma\pm2\sigma$&mean$\pm1\sigma\pm2\sigma$&mean$\pm1\sigma\pm2\sigma$\\
  \hline
$\Omega_{de}$ & 0.692$_{-  0.012}^{+  0.012}$ $_{-  0.023}^{+  0.022}$&0.692$_{-  0.011}^{+  0.012}$ $_{-  0.023}^{+  0.022}$&0.691$_{-  0.012}^{+  0.012}$ $_{-  0.023}^{+  0.022}$\\
$w_0$ &-1.023$_{-  0.058}^{+  0.061}$ $_{-  0.126}^{+  0.121}$&-1.025$_{-  0.057}^{+  0.062}$ $_{-  0.126}^{+  0.119}$&-1.025$_{-  0.058}^{+  0.063}$ $_{-  0.126}^{+  0.121}$\\
$c_a^2$ &-1.465$_{-  0.535}^{+  0.147}$ $_{-  0.535}^{+  0.774}$&-1.458$_{-  0.542}^{+  0.147}$ $_{-  0.542}^{+  0.809}$&-1.481$_{-  0.519}^{+  0.141}$ $_{-  0.519}^{+  0.755}$\\
$r$ & 0.025$_{-  0.025}^{+  0.003}$ $_{-  0.025}^{+  0.056}$&0.026$_{-  0.026}^{+  0.006}$ $_{-  0.026}^{+  0.041}$&0.019$_{-  0.019}^{+  0.004}$ $_{-  0.019}^{+  0.035}$\\
$n_t$ &-0.242$_{-  0.050}^{+  0.242}$ $_{-  0.441}^{+  0.242}$&-0.224$_{-  0.043}^{+  0.224}$ $_{-  0.423}^{+  0.224}$&-0.251$_{-  0.054}^{+  0.251}$ $_{-  0.450}^{+  0.251}$\\
$10\Omega_bh^2$ & 0.223$_{-  0.002}^{+  0.002}$ $_{-  0.003}^{+  0.003}$&0.222$_{-  0.002}^{+  0.002}$ $_{-  0.003}^{+  0.003}$&0.222$_{-  0.002}^{+  0.001}$ $_{-  0.003}^{+  0.003}$\\
$\Omega_{cdm}h^2$ & 0.119$_{-  0.001}^{+  0.001}$ $_{-  0.003}^{+  0.003}$&0.119$_{-  0.001}^{+  0.001}$ $_{-  0.003}^{+  0.003}$&0.119$_{-  0.001}^{+  0.001}$ $_{-  0.003}^{+  0.003}$\\
$h$ & 0.679$_{-  0.012}^{+  0.011}$ $_{-  0.023}^{+  0.024}$&0.679$_{-  0.012}^{+  0.011}$ $_{-  0.023}^{+  0.024}$&0.679$_{-  0.012}^{+  0.011}$ $_{-  0.023}^{+  0.024}$\\
$n_s$ & 0.966$_{-  0.005}^{+  0.005}$ $_{-  0.009}^{+  0.009}$&0.966$_{-  0.005}^{+  0.005}$ $_{-  0.009}^{+  0.009}$&0.966$_{-  0.005}^{+  0.005}$ $_{-  0.009}^{+  0.009}$\\
$\log(10^{10}A_s)$ & 3.056$_{-  0.025}^{+  0.024}$ $_{-  0.049}^{+  0.049}$&3.056$_{-  0.025}^{+  0.025}$ $_{-  0.049}^{+  0.049}$&3.059$_{-  0.024}^{+  0.024}$ $_{-  0.049}^{+  0.049}$\\
$\tau_{rei}$ & 0.061$_{-  0.013}^{+  0.013}$ $_{-  0.027}^{+  0.027}$&0.062$_{-  0.013}^{+  0.013}$ $_{-  0.026}^{+  0.027}$&0.063$_{-  0.013}^{+  0.013}$ $_{-  0.026}^{+  0.027}$\\
\end{tabular}\label{tab}
\end{table}

\begin{figure}[htb]
\centerline{
\includegraphics[width=0.5\textwidth]{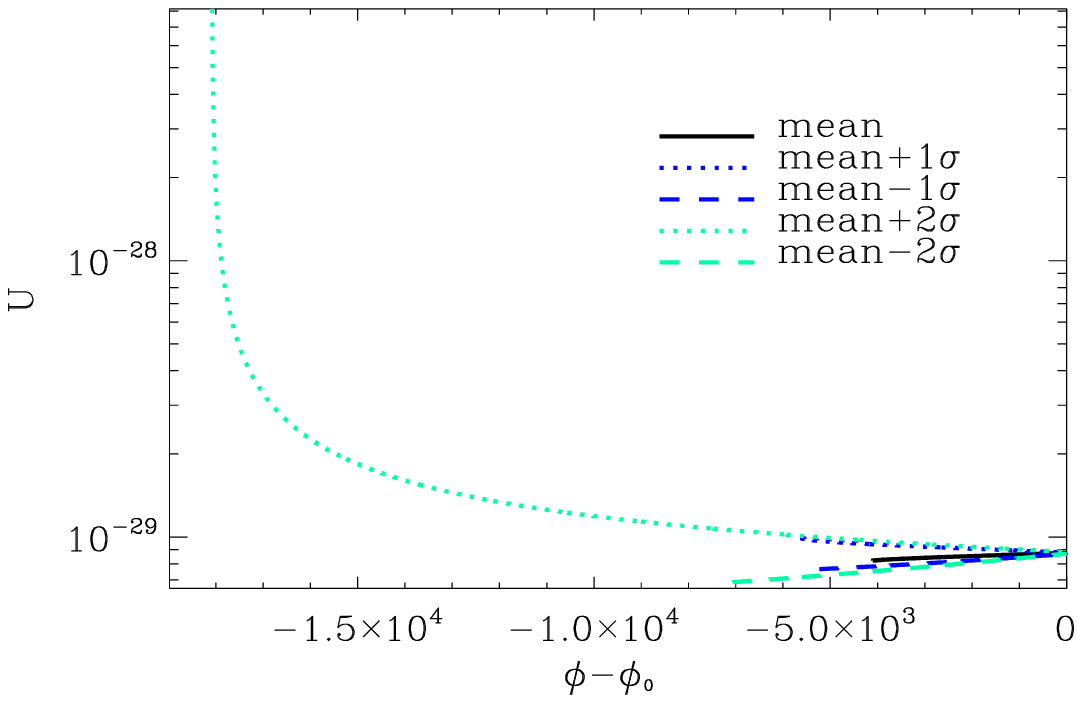}\includegraphics[width=0.5\textwidth]{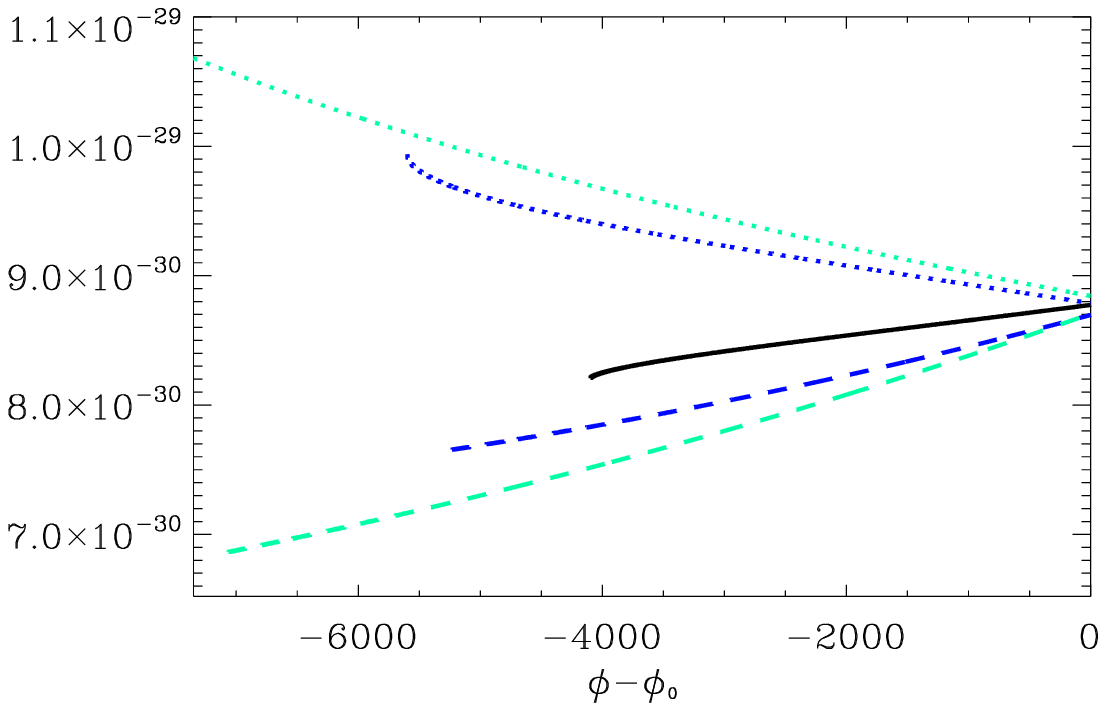}}
\caption{The reconstructed potentials for models corresponding to the mean values as well as the upper and lower 1$\sigma$ and 2$\sigma$ limits of parameters from Table \ref{tab}. Left: reconstruction for $a$ from 0.001 to 1 (current epoch) for each potential. Right: enlarged part of the left plot.}
\label{uphi}
\end{figure}

Reconstructed potentials of the dark energy scalar field are shown in Fig. \ref{uphi} (more details on reconstruction of the potentials for fields with $c_a^2=0$ can be found in \cite{novosyadlyj2009}, for quintessential fields with arbitrary $c_a^2$ in \cite{sergijenko2011}; the impact of uncertainties in the cosmological parameters determination on the reconstructed potentials for $w=const$ was analyzed in \cite{sergijenko2008} and for $c_s^2=const$ was estimated in \cite{sergijenko2015}). We see the phantom fields slowly rolling up the potentials for mean values of parameters, their 1$\sigma$ and 2$\sigma$ lower confidence limits, as well as the quintessence fields slowly rolling down the potentials for 1$\sigma$ and 2$\sigma$ upper confidence limits during the evolution of the Universe from $a=0.001$ to 1.

This work was supported by the project of Ministry of Education and Science of Ukraine (state registration number 0116U001544). Author also acknowledges the usage of CosmoMC package.

\end{document}